\documentclass{article}

\usepackage{graphicx}  
\usepackage{amsmath}   
\usepackage[compress]{cite}
\usepackage{amssymb}   
\usepackage{bm} 
\usepackage{mathrsfs}

\def\eq#1{{Eq.~(\ref{#1})}}
\def\eqs#1{{Eqs.~(\ref{#1})}}

 \def\g{\sqrt{-g}\,}

\def\half{{\frac{1}{2}}}

\newcommand{\df}{\delta}

\def\cc{{cosmological\ constant}}

\newcommand{\LL}{Lanczos-Lovelock }

\def \lp { L_0 }
\def\l{\left}
\def\r{\right}
\def \mA {A\l[\sigma;\lp \r]}

 \title{Emergent Gravity Paradigm: Recent Progress}

\author{T. Padmanabhan\\
IUCAA, Pune University Campus,\\
  Ganeshkhind, Pune- 411 007.\\
{\small{email: paddy@iucaa.ernet.in}}
}
\date{ }

\begin{document}

\maketitle

\begin{abstract}
 Research during the last one decade or so suggests that the gravitational field equations in a large class of theories (including, but not limited to, general relativity) have the same status as the equations of, say, gas dynamics or elasticity.  This paradigm  provides  a refreshingly different way of interpreting spacetime dynamics and highlights the fact that several  features of classical gravitational theories have direct thermodynamic interpretation. I review the recent progress in this approach, achieved during the last few years.  
\end{abstract}

  \section{Introduction and Summary}

In the study of fluid mechanics or gas dynamics, we treat the system as a continuum and use variables like density, fluid velocity, pressure, etc. which can be defined through purely \textit{mechanical} considerations. But these  are not sufficient and they need to be supplemented by \textit{thermodynamic} variables  like temperature, entropy etc., which, however, cannot be really understood within the continuum limit. 
Boltzmann was the first to emphasize that a consistent description of matter requires interpreting the temperature in terms of the energy stored in the \textit{discrete}  microscopic degrees of freedom through the principle:  ``If you can heat it, it must have microstructure''. Boltzmann used this idea successfully to \textit{infer} the discrete, atomic nature of matter from the thermal phenomena,  much before we really understood what the atoms are. The equations of, say, fluid mechanics then emerge in the continuum limit of the statistical mechanics applied to the discrete degrees of freedom. 

Since any spacetime (just like normal matter) will be perceived to be hot by some observers, the Boltzmann principle could again be applied to interpret the continuum physics of the spacetime as a thermodynamic description of unknown microscopic degrees of freedom (``atoms of spacetime''). The emergent gravity paradigm, as described here,  is such an attempt to obtain and interpret the \textit{field equations} of gravitational theories in a thermodynamic context.
The crucial thermodynamic inputs, viz. the temperature and entropy density of spacetime, are provided by the Davies-Unruh temperature \cite{du} (and a corresponding entropy) attributed to the null surfaces which are perceived as horizons by local Rindler observers.
\textit{It turns out that, using this single quantum input, one could rephrase and re-derive the entire description of classical gravity in a novel language}. 

More specifically, investigations along these lines have led to the following results in the recent years: 

\begin{enumerate}

 \item 
This paradigm provides a direct link between certain \textit{dynamical} variables (built from the metric and Christoffel symbols) and the \textit{thermodynamical}
variables like temperature and entropy \cite{KBP}. 
 It also provides a  thermodynamic interpretation for the Noether current (and charge) associated with the time translation vector\cite{grtp}  in any \textit{arbitrary} spacetime. In addition to providing the foundation for interpreting spacetime dynamics as thermodynamics, these results also  help us to understand\cite{grtp} several features  in general relativity like, e.g., the  age-old problem of the factor of two  (see e.g., Ref. \cite{grtp28})  in defining the mass and angular momentum using Killing vectors. 

\item
It is possible to \textit{derive} the field equations of a large class of gravitational theories (including  general relativity) from a thermodynamic extremum principle \cite{aseemtp1,aseemtp2}. Further, this  extremum principle remains invariant under the transformation $T_{ab} \to T_{ab} + \rho_0 g_{ab}$ thereby \textit{making gravity immune to shifts in the zero level of the energy density}. This immunity of gravity to the vacuum energy density is an important fact about the dynamics of gravity\cite{tpap} which is not incorporated in the usual approaches to gravity that treat the metric as a dynamical variable. 

\item 
The resulting field equations (which are equivalent to the standard equations, with the cosmological constant arising as an integration constant) have an elegant interpretation in terms of suitably defined bulk ($N_{\rm bulk}$) and surface ($N_{\rm sur}$) degrees of freedom in a region of 3-space. It turns out\cite{grtp} that all static spacetimes obey the condition of \textit{holographic equipartition}, viz. $N_{\rm bulk} = N_{\rm sur}$. More generally, the time evolution of the metric is driven by the holographic discrepancy ($  N_{\rm sur} - N_{\rm bulk}$) between the surface and bulk degrees of freedom.

\item 
This approach suggests that there exists a new conserved quantity for our universe  (``CosMIn'';  see Refs.~\cite{hptp1,hptp2}) which should have the value $4\pi$. Using this, one can relate the numerical value of the cosmological constant --- which arises as an undetermined integration constant to the field equations in this approach --- to two other parameters of high energy physics. Remarkably enough, this leads\cite{hptp1,hptp2} to the observed value of the cosmological constant in our universe.

\item 
These results continue to hold in all \LL\ models of gravity which are much more general than general relativity \cite{llreview}. \textit{In particular, the horizon entropy in \LL\ models is not proportional to the area} and hence it is a nontrivial fact that  the results extend  to these models. This  suggests that the emergent paradigm transcends general relativity  and is telling us something deeper about the nature of spacetime. 

\item 
Proceeding from the thermodynamic limit to the statistical mechanics, it is  possible to relate
\cite{cheshire1,cheshire2} the  thermodynamic extremum principle, which is   used to derive the field equations ---
 mentioned in item (2) above   --- to the zero-point-length of spacetime \cite{zpl} arising from the quantum discreteness at Planck scales. This identifies the essential features
 of the microscopic theory    which could lead to the thermodynamic description of gravity in the long wavelength limit. 

\end{enumerate}

 I will now elaborate on each of these aspects, concentrating on the recent work leading to the above results. 
Other approaches, which are somewhat similar in spirit but do not contribute to the understanding of above features,  are  not discussed here. 
Further, I use the phrase `emergent gravity' to imply that the \textit{field equations} have an emergent interpretation, rather than  speculate whether the spacetime, its manifold structure, etc. are themselves emergent \cite{carlip}.
An earlier review of mine  \cite{insights} has a broader discussion and a more extensive set of references. 

I will use the `mostly positive' signature and units with $\hbar=c=16\pi G=1$ in most places so that the field equations in general relativity  read $2G_{ab}=T_{ab}$. The English alphabets range over $0,1,2,..d=D-1$ while the Greek  alphabets  range over spatial coordinates $1,2,..d=D-1$. Most of the discussion will be confined to $D=4,d=3$. The occurrence of $\equiv$ in an equation indicates that the equation defines a particular variable.

\section{Kinematics of spacetime geometry and its thermodynamic interpretation}

Judicious application of the principle of equivalence and principle of general covariance suggests that the effect of gravity on matter  can be described --- quite elegantly --- by interpreting the gravitational field as due to spacetime curvature. Starting with this premise, we will consider a spacetime  in some arbitrary ($1+d$) foliation based on a time function $t(x^a)$, with a unit normal $u_a(x^i)\propto \nabla_at$. This will split the metric $g_{ab}$ into the usual components, viz. the lapse ($N$), shift ($N_\alpha$) and $d$-metric $h_{ab} = g_{ab} + u_au_b$. The foliation also introduces the  extrinsic curvature $K_{ab}$ of the $t=$ constant surfaces, and the useful  combination $p_{ab} \equiv K_{ab} - h_{ab}K$.
Treating $u^a$ as the
four-velocity of the congruence of observers leads to the  (in general, nonzero) acceleration vector $a_i\equiv u^j\nabla_ju_i=h^j_i(\nabla_jN/N)$ which is purely spatial (i.e., $u^ia_i=0$) and has the magnitude $a\equiv\sqrt{a_ia^i}$.

The conditions $t(x)=$ constant, $N(x)=$ constant, taken together, define the codimension-2 surface $\mathcal{S}$ (which is a natural generalization of the equipotential surface) with the area element $\sqrt{\sigma}d^{D-2}x$ and the binormal $\epsilon_{ab}\equiv q_{[a}u_{b]}$ where
$q_i\propto \nabla_iN$ is the unit normal to the $N(x)=$ constant surface.
 Since only the component $h^i_jq_j$ of $q_i$ which is normal to $u_i$ contributes to this binormal, we can take it to be $\epsilon_{ab}\equiv r_{[a}u_{b]}$  where  $r^\alpha=\epsilon(a^\alpha/a)$ is essentially the unit vector along the acceleration.
 (The factor $\epsilon=\pm1$ ensures that the normal $r^\alpha$ is pointing outwards irrespective of the direction of acceleration. We will usually assume $\epsilon=1$, when this distinction is not important.) 

Comoving observers with $x^\alpha=$ constant will have the acceleration $a$ which allows us to introduce the notion of a local Rindler frame at any event along the following lines \cite{a12} (We will work with $D=4$ for simplicity.): We first introduce the local inertial frame $(T, \mathbf{X})$ in a region around any event $\mathcal{P}$, and align, say, the $X$-axis along the direction of the acceleration in the original frame. (The special case when the comoving observers have zero acceleration can be handled as a limiting case; we will discuss it briefly later on.) We next boost from the inertial frame to a Rindler frame $(t,\mathbf{x})$ with acceleration $a$ using the standard transformations: $X=x\cosh (at), T=x\sinh (at)$. A null surface passing though $\mathcal{P}$, which gets mapped to the $X=T$ surface in the local region, where the inertial frame is introduced, will now act as a patch of horizon to the $x=$ constant Rindler observers. They will attribute the (Tolman-corrected) temperature $T=Na/2\pi$ to the vacuum state of the freely falling observers. Further,  one quarter of the area element $dS=\sqrt{\sigma}d^2x/4 L_P^2$ can be thought of as the entropy associated with this patch of horizon \textit{in general relativity}. (We will discuss the more general situation later on). This result introduces a quantum of area $L_P^2$; the factor (1/4) is then purely conventional, because one could have as well worked with $4L_P^2$ rather than $L_P^2$. All these can be introduced  \textit{purely kinematically} by studying quantum field theory in a  pre-specified curved spacetime, without  introducing any dynamics for the gravitational field. 

  As we shall soon see, the thermodynamic description of spacetime geometry becomes most apparent \cite{KBP} if we use the variables 
\begin{equation}
f^{ab} \equiv \g g^{ab} ; \qquad N^{a}_{bc} \equiv -\Gamma^{a}_{bc}+\frac{1}{2}(\Gamma^d_{bd}\delta^{a}_{c}+\Gamma^d_{cd}\delta^{a}_{b}) 
\label{NGamma} 
\end{equation} 
instead of the standard pair ($g_{ab}, \Gamma^i_{jk}$). It turns out that these variables $(f^{ab},N^i_{jk})$ --- or more precisely  their variations --- have a direct thermodynamic significance in the following manner:
Let $\mathcal{H}$ be  a null surface with temperature $T$ and entropy density $s=\sqrt{\sigma}/4$ attributed to it by local Rindler observers who perceive it as a horizon. (It is convenient to use the entropy per unit \textit{coordinate} area $dS/d^2x=\sqrt{\sigma}/4$ in what follows; one could, of course, translate everything into entropy per unit \textit{proper} area $dS/\sqrt{\sigma}d^2x=1/4$ if one wants. This does not change anything.)
Then, \cite{KBP} we can show that (in units with $G=L_P^2$):

\begin{itemize}
 \item The integral of $ N^{c}_{ab} f^{ab}$  over $\mathcal{H}$ can be interpreted \cite{a41} as its heat content $Ts$; that is:
 \begin{equation}
  \frac{1}{16 \pi L_P^2}\int d^3 \Sigma_c (N^{c}_{ab} f^{ab}) =\int d\lambda\ d^2x\ T  s
 \end{equation} 
 
 \item More remarkably, the variations $f\delta N$ and $N\delta f$ possess corresponding thermodynamic interpretations\cite{KBP} for  variations which preserve the null surface:
 \begin{eqnarray}
\frac{1}{16 \pi L_P^2}\int  d^3 \Sigma_c(N^{c}_{ab}\df f^{ab})&=&  \int d\lambda\ d^2x\ T \df s; 
\label{stsdt0}\\
\frac{1}{16 \pi L_P^2}\int  d^3 \Sigma_c (f^{ab}\df N^{c}_{ab})&=&  \int d\lambda\ d^2x\ s \df T
\label{stSdT}
\end{eqnarray}
We thus see that  the variations ($N\delta f, f\delta N$) show \textit{thermodynamic conjugacy} similar to the corresponding   $(T\delta s, s\delta T)$. Of these, we can think of $f^{ab}$ as an extensive variable and $N^i_{jk}$ as an intensive variable, just as in conventional thermodynamics.
\end{itemize}

The complementary nature of $SdT$ and $TdS$ in \eqs{stsdt0} and (\ref{stSdT}) is often ignored in literature and hence is worth emphasizing. For example, the Schwarzschild black hole (with horizon area $A$ and $S=4\pi M^2, E=M,T=1/8\pi M$) satisfies the relation 
\begin{equation}
E = 2TS = \frac{1}{2} \frac{A}{L_P^2}T =\frac{1}{2} N_{sur} T 
\end{equation} 
with  crucial factors of 2 [and $(1/2)$] in these equations. So if we attribute $N_{\rm sur} = A/L_P^2$ degrees of freedom to the horizon area $A$, then each degree of freedom carries  $(1/2) k_BT$ amount of energy.
 The often quoted relation $\delta E = T\delta S$, when  we add an amount of energy $\delta E=\delta M$ --- though algebraically correct --- is conceptually misleading because it suggests that  $T$ is kept constant while the process takes place. This, of course, is not true since   both $T$ and $S$ change when $M$ changes. What we actually have is
$
\delta E = 2S\delta T + 2T \delta S  
$
with an additional relation $S \propto M^2 \propto T^{-2}$ which is maintained during the variation.  This   allows us to express $\delta E$ either with $\delta S$ or with  $\delta T$ alone:
\begin{equation}
   \delta E = T\delta S = - 2S \delta T = - \frac{1}{2} \frac{A}{L_P^2} \delta T = - \frac{1}{2} N_{\rm sur} \delta T
\label{e4}
  \end{equation} 
 So, the addition of energy can \textit{also} be thought of as resulting in an increase in the temperature with $N_{\rm sur}$  held fixed, with the  minus sign indicating the negative specific heat of the gravitating system! 
  In the general context of null surfaces and other boundaries, we will find that it is the  interpretation involving $S\delta T$ (which corresponds to $f^{bc}\delta N^a_{bc}$) that provides a more natural description (and is generally covariant).

These variables $(f^{ab},N^i_{jk})$ also appear in the  conserved currents associated with vector fields in the spacetime. It is rather trivial to obtain a conserved current $J^a$ from \textit{any} vector field $v^a$ in the spacetime by the following procedure. 
If we separate the derivative $\nabla^l v^m$ of  any vector field $v^j$   into the symmetric and anti-symmetric parts by 
$\nabla^{(l} v^{m)} \equiv S^{lm}$ and $\nabla^{[l} v^{m]} \equiv J^{lm}$, then   $J^{lm}$  immediately gives us a conserved current $J^i\equiv \nabla_k J^{ik}$.   A more useful form for $J^a[v]$ can be found as follows:
Using the Lie derivative of the connection
$
\pounds_v\Gamma^a_{bc}=\nabla_b \nabla_c v^a+R^a_{\phantom{a}cmb}v^m
$,
in \eq{NGamma}, we get the relation: 
$
g^{bc}\pounds_vN^a_{bc} =\nabla_bJ^{ab}-2R^a_bv^b.
$
This leads to  an  explicit form of the conserved current:
\begin{equation}
 J^a[v] = \nabla_b J^{ab} [v] = 2 R^a_b v^b + g^{ij} \pounds_v N^a_{ij}
\label{noe}
\end{equation}
In fact, this \textit{is} indeed the usual Noether current\footnote{The overall proportionality constant in any conserved current is arbitrary and the above expression will give the usual Noether current in units with $16\pi G=1$; when we switch to $G=1$ units, the left hand side of \eq{noe} should be multiplied by $16\pi$ to get the standard expressions.} associated with  $v^a$ --- which we have now obtained purely kinematically \textit{without mentioning the action principle for gravity or any diffeomorphism invariance!}. 

While \eq{noe} associates a conserved current $J^a[v]$ with \textit{any} vector field $v^a$, the current related to the  time evolution vector, $\xi^a \equiv N u^a$, is of special interest. This vector  measures the  proper-time lapse corresponding to the normal $u_a  = - N \nabla_a t$ to the  $t=$ constant surfaces. (In static spacetimes,   $\xi^a$ can be chosen to be the timelike Killing vector.)   An elementary calculation shows \cite{grtp} that the Noether potential and charge associated with $\xi^a$ have  simple forms which --- as we shall see --- admit a direct thermodynamic interpretation. We find that (in units with $G=1$)
\begin{equation}
 \sqrt{\sigma}J_{ab} = [T_{loc}s]\epsilon_{ab}, \quad  \sqrt{\sigma}J_{ab}u^a =  [T_{loc}s]r_b, 
\quad \sqrt{\sigma}J_{ab}r^a =[T_{loc}s]u_b
\end{equation} 
 where $\epsilon_{ab}= r_{[a}u_{b]}$ is the binormal to the $t=$ constant, $N(t,\mathbf{x})=$ constant surface with area element $\sqrt{\sigma}d^2x$. The second and third relations give currents of the heat (enthalpy) density $h\equiv T_{loc}s$. 
 More importantly, we can show that:
\begin{equation}
u_aJ^a(\xi)=\frac{1}{4} D_\alpha (T_{\rm loc} r^\alpha)
\label{fin1}
\end{equation} 
where  $D_\alpha$ is the covariant derivative on the $t=$ constant surface. 
Integrating \eq{fin1}  over $\sqrt{h}d^3x$, we obtain the total Noether charge contained inside a volume. In particular, if we choose the region to be bounded by the equipotential surface
$N(t,\mathbf{x})=$  constant surface within the $t=$ constant surface, we get \cite{grtp}:
\begin{equation}
2\int_\mathcal{V}\sqrt{h}\, d^3x\ u_aJ^a[\xi]=
\epsilon\int_{\partial\mathcal{V}}\frac{\sqrt{\sigma}\, d^2x}{L_P^2} \left(\frac{1}{2}T_{\rm loc}\right)
\label{ib1}
\end{equation}
where we have re-introduced $G=L_P^2$.
This equation relates (twice) the Noether charge contained in $\mathcal{V}$  to the  equipartition energy of the surface,  attributing one  degree of freedom to each cell of Planck area $L_P^2$.
Alternatively, if we think of
$s=\sqrt{\sigma}/4L_P^2$ as the analogue of the entropy density,  we  get: 
\begin{equation}
\int_\mathcal{V}\sqrt{h}\ d^3x\ u_aJ^a[\xi]
=\epsilon\int_{\partial\mathcal{V}}d^2x\ Ts
\label{flux3}
\end{equation}
which relates the Noether charge to  the heat (enthalpy)  density ($TS/A$) of the boundary surface. 
 This delightfully simple interpretation  of the Noether charge is valid in \textit{the most general} context without any assumptions like static nature, existence of Killing vectors, asymptotic behaviour, etc.

Incidentally, the
 factor 2 on the left hand side of \eq{ib1} solves  an old puzzle\cite{grtp28} known to general relativists. The integral on the right hand side of \eq{ib1} gives $(1/2)TA=2TS$ if we assume (for the sake of illustration) that $T$ = constant on the boundary and $S=A/4$. Then, the Noether charge $Q$ is just the heat content  $Q=TS$. Therefore, the Noether charge is \textit{half} of the  equipartition energy of the surface $(1/2)TA=2TS$ if we attribute $(1/2)T$ to each surface degree of freedom. For example, in the case of the Schwarzschild geometry,  the   equipartition energy  is equal to the total mass $M=2TS$. \textit{But what the Noether charge measures is not the energy $E$ but the heat content  $E-F=TS$ which has half this value, viz. $(M/2)$.} This  leads to a ``problem''  in standard general relativity, when one tries to define the total mass of a spacetime (which behaves like the Schwarzschild spacetime asymptotically) using the  Komar integral. In this calculation, $\xi^a$ will be taken to be  the standard timelike Killing vector and the Noether potential will become the Komar potential. The integral one calculates using  the Killing vector $\xi^a$ is identical to the one in the  computation of the Noether charge  and the answer, of course, will be  $TS = (M/2)$. In standard general relativity, this was considered  puzzling\cite{grtp28} because, in that context, we  only have a notion of energy but no notion of heat content ($TS$), free energy ($F=E-TS$), etc. The thermodynamic  paradigm  --- which introduces the $\hbar$ through the definition of the Davies-Unruh temperature $k_BT=(\hbar/c)(\kappa/2\pi)$ from an acceleration $\kappa$ --- shows that the Noether charge is the heat content (enthalpy) $TS$ and \textit{not} the energy $2TS$, and that the result \textit{must} be $M/2$ for consistency of the formalism.

In other words, the standard approach to general relativity  can only interpret $M$ physically (as energy), while the thermodynamic approach allows us to \textit{also} interpret $M/2$ physically as the heat content $TS$. \textit{This is an example of the emergent paradigm giving us  a deeper insight into some puzzling features of standard general relativity.}

The key role played by the vector field $g^{ij}\pounds_\xi N^a_{ij}$ in the expression for the Noether current is noteworthy. This combination, which  has  a direct thermodynamic interpretation,  will continue to play an important role throughout our discussions.  
One can also show that 
\begin{equation}
 \sqrt{h} u_a g^{ij} \pounds_\xi N^a_{ij}=-h_{ab}\pounds_\xi p^{ab}; 
\quad p^{ab}\equiv\sqrt{h}(Kh^{ab}-K^{ab})
\label{niskp}
\end{equation}
which provides a direct relationship with the extrinsic curvature of the foliation. 
In fact, it is possible to obtain the heat density of the local Rindler horizon more directly from the extrinsic curvature tensor along the following lines. The  expression
 \begin{equation}
  H_{\rm sur}\equiv\frac{\partial }{\partial t}\left[\frac{1}{8\pi  L_P^2}\int_{\mathcal{H}} K\sqrt{h}dtd^2x\right]
  \end{equation} 
when evaluated on a local Rindler horizon $\mathcal{H}$ with surface gravity $\kappa$ and transverse area $A_\perp$, gives \cite{a41}
the heat content:
\begin{equation}
H_{\rm sur}\to \frac{\kappa A_\perp}{8\pi  L_P^2}=TS                                             
\end{equation} 
 If we now analytically continue the local Rindler time coordinate $t$ to the Euclidean sector, then the natural range of integration for the Euclidean time $t_E$ is $0<t_E<(2\pi/\kappa)$. This will give the entropy:
 \begin{eqnarray}
  \int_{\mathcal{H}} dt_Ed^2x K\sqrt{h} 
  = \frac{2\pi}{\kappa}\times\left(\frac{\kappa A_\perp}{8\pi  L_P^2} \right)=\frac{A_\perp}{4L_P^2}=S
  \label{entro}
 \end{eqnarray}  
So the entropy \textit{density} of spacetime, when evaluated around any event after the Euclidean continuation, is also equal to $(K\sqrt{h})$ in the local Rindler approximation.
We will need this result later on.

For future reference, we note that the Noether potential  and
the entropy density of a patch of a local Rindler horizon can also be written in the form: 
\begin{eqnarray}
 J^{ab}&=&\frac{1}{8\pi}P^{abcd}\nabla_c\xi_d= \frac{1}{4}  T_{loc}\left(P^{abcd}\epsilon_{cd}\right);\nonumber\\
s&=&-\frac{\sqrt{\sigma}}{2T_{loc}}J^{ab}\epsilon_{ab}=-\frac{1}{8}\sqrt{\sigma} P^{abcd} \epsilon_{ab}\epsilon_{cd}  
\label{introP}                                                                                                  
\end{eqnarray} 
where $P^{abcd}$ is an `entropy tensor' with the following properties: (a) it is made from the metric tensor; (b) it has all the algebraic symmetries of the curvature tensor; (c) it is divergence free in all indices. In the above discussion $P^{ab}_{cd}=(1/2)(\delta^a_c\delta^b_d-\delta^b_c\delta^a_d)$ is proportional to the determinant tensor but we will see later that the expressions in \eq{introP} have a far greater domain of validity.

\section{Dynamics of spacetime geometry from a thermodynamic variational principle}

The discussion so far has been completely kinematical in the sense that we have treated the metric $g_{ab}$ as pre-assigned. We will now turn to the question of how  matter curves the spacetime.  From a thermodynamic perspective, one would like to obtain this result by extremizing a suitable thermodynamic potential which --- based on the results of the previous sections --- could be  the heat density of the null surfaces which act as local Rindler horizons. I will now describe how this can be done and how it incorporates a key dynamical principle about gravity which is  ignored in conventional approaches.

\subsection{The single most important fact about the gravitational dynamics}\label{sec:6.1}

To motivate this approach, I begin by stressing the \textit{single most important fact} about gravitational dynamics\cite{grtp} which ---  because of a historical accident --- is  completely ignored in the usual approach: \textit{Gravity does not couple to  the bulk energy density arising from the addition of a constant to the  matter Lagrangian}. Any attempt to describe the dynamics of gravity without incorporating this observed feature is bound to be wrong.

This fact demands that the  gravitational field equations \textit{must be} invariant under \textit{the symmetry transformation of the matter sector} equations:
\begin{equation}
L_{\rm matter} \to L_{\rm matter}+ \mathrm{constant},
\label{lmsym}
\end{equation}  
resulting in $T^a_b \to T^a_b + ({\rm constant})\ \delta^a_b$. (For example, the electroweak symmetry breaking is equivalent to  the shifting of the standard model Lagrangian by a large constant and we know that the evolution of the universe was  not affected by this transition.) 
Ensuring this symmetry for gravitational dynamics will be conceptually nice because all other (non-supersymmetric) theories have field equations which remain invariant if we add a constant to the Lagrangian.

The usual gravitational field equations, in contrast to matter field equations, are of course \textit{not} invariant under the addition of a constant to the matter Lagrangian. This addition
 changes the energy-momentum tensor of the matter by 
$
 T^a_b \to T^a_b + ({\rm constant}) \ \delta^a_b
$ 
and the usual gravitational field equations now become $\mathcal{G}^a_b = T^a_b + ({\rm constant}) \ \delta^a_b$ which is equivalent to the introduction of a \cc\ (if $\Lambda =0$ originally) or changing its numerical value, if $\Lambda  \ne 0$  in the original gravitational Lagrangian. This  \textit{is} the key problem related to the \cc, viz., that its numerical value (either zero or non-zero) can be altered by the transformation in \eq{lmsym} \textit{which leaves the matter equations of motion unchanged}. A particle physicist  working with  the standard model can choose the overall constant in the matter Lagrangian arbitrarily because the standard model does not care for this constant. But each choice for this constant will lead to a different value for the cosmological constant and a different geometry for the universe, many of which will turn out to be observationally untenable.

An alternative way of stating this problem is as follows: Suppose we discover a fundamental principle that helps us to determine the numerical value of the cosmological constant (either zero or non-zero). Such a principle is quite useless if the gravitational field equations are not invariant under the transformations in \eq{lmsym}.

This discussion helps us to  identify three  ingredients which are \textit{necessary} to solve the \cc\ problem: 
\begin{enumerate}
 \item The gravitational field equations \textit{must be}  invariant under the transformations in \eq{lmsym}  so that gravity is immune to the shift in the zero level of the energy densities.
 
 \item At the same time, the \textit{solutions} to the gravitational  field equations \textit{must} allow the cosmological constant to influence the geometry of the universe, because this is needed to explain the observed accelerated expansion of the universe. 
 
 \item We also require a fundamental physical principle to determine the numerical value of the \cc\ since it cannot be introduced as a low energy parameter in the Lagrangian if the theory is invariant under the transformation in \eq{lmsym}.
\end{enumerate}

The first two requirements above might  sound impossible to satisfy simultaneously, but it can be done! The trick is to keep the gravitational field equations  invariant under the transformation $T^a_b\to T^a_b+$(constant)$\delta^a_b$  but allow the inclusion of a \cc\
as an \textit{integration constant} in the solutions. Consider, for example,  a theory in which the field equations are given by the requirement that:
\begin{equation}
 (\mathcal{G}^a_b - T^a_b) \ell_a \ell^b = 0
 \label{tf}
\end{equation} 
for all \textit{null} vectors $\ell^a$ in the spacetime \cite{aseemtp1,aseemtp2}. 
In \eq{tf},  $\mathcal{G}^a_b = 2 G^a_b$ in general relativity and could be some other  tensor in alternate theories of gravity, but necessarily satisfying the generalized Bianchi identity $\nabla_a \mathcal{G}^a_b=0$.
Equation (\ref{tf}) can be solved  by $\mathcal{G}^a_b - T^a_b = F(x) \delta^a_b$, but the generalized  Bianchi identity
($\nabla_a \mathcal{G}^a_b =0$)
and the conservation of the energy-momentum tensor $(\nabla_a T^a_b =0)$ imply that $F(x)$ must be a constant. Therefore, \eq{tf} is actually equivalent to $ \mathcal{G}^a_b = T^a_b + \Lambda \delta^a_b$ with an arbitrary \cc\ $\Lambda$ appearing as an integration constant. Thus  a theory of gravity in which the field equations reduce to those in \eq{tf}, will satisfy the first two requirements in our list for solving the \cc\ problem. (We will address the third requirement later on.)

This turns out to be a very strong demand and has important consequences, usually overlooked in attempts to ``solve'' the \cc\
problem. To see this, consider any theory of gravity interacting with matter that satisfies the following  conditions: 
\begin{enumerate}
 \item The theory is generally covariant  and the matter action is obtained by integrating  a scalar Lagrangian $L_m(g_{ab},\phi_A)$ over the measure $\sqrt{-g} d^4x$. 
 \item The  equations of motion for the matter sector are invariant under the transformation $L\to L + C$ where $C$ is a scalar constant.
 \item The gravitational field equations are obtained by  the  variation of the metric tensor $g_{ab}$ 
in an \textit{unrestricted} manner
in the total action (which is obtained by integrating a local Lagrangian over the  spacetime).
\end{enumerate}
We \textit{cannot} solve the \cc\ problem in any theory satisfying the above three requirements.\cite{tpap} It follows that  one cannot obtain the gravitational field equations of the form in \eq{tf} in any theory which satisfies the above three criteria.

So, even though all the three criteria given above seem very reasonable, they together will prevent us from solving the \cc\ problem and we must  give up at least one of them. If we do not want to give up locality, general covariance  or the freedom to add a constant to the matter Lagrangian, we can only tinker with the third requirement.\footnote{One can  obtain \eq{tf} if we postulate that the gravitational field equations are obtained by varying the metric but keeping $\sqrt{-g} = $ constant. Such  unimodular theories of gravity 
--- which  bypass the condition (3) above ---
 have been studied in the literature in the past \cite{unimod1,unimod3,unimod4}. Unfortunately, the motivation to keep $\sqrt{-g} = $ constant is quite weak.} 
I will now show how  \eq{tf} arises naturally if we use a thermodynamic extremum principle. 

\subsection{Gravitational field equations from a thermodynamic extremum principle}

In this approach, we will associate a heat density with all null surfaces in the spacetime. Maximization of this heat density for \textit{all} null surfaces simultaneously will then lead to \eq{tf}. Let $\ell_a$ be a null congruence defining a null surface which is affinely parametrized. 
Then we extremize the expression 
\begin{equation}
 Q=\int_{\lambda_1}^{\lambda_2} d\lambda\ d^2x\, \sqrt{\sigma}\, \left[h_g(\ell) + h_{matt}  \right]= \int_{\lambda_1}^{\lambda_2} d\lambda\ d^2x\, \sqrt{\sigma}\, \left[h_g(\ell) +  T_{ab}\ell^a\ell^b \right],
\end{equation}
(where $h_{matt}=T_{ab}\ell^a\ell^b$ and $h_g(\ell)$ are the heat densities of matter and gravity) over all $\ell_a$ simultaneously. The condition that $Q$ is an extremum for all null surfaces (or null vectors $\ell^a$) leads to a constraint on the background metric which will be equivalent to the field equations in the form of \eq{tf}. This maximization involves varying the null vector fields rather than the metric and hence it bypasses  the third requirement in our earlier list.
 In such an approach,  the  variational principle  itself (not just the field equations) is invariant under the transformation $T^a_b\to T^a_b+$(constant)$\delta^a_b$.

The resulting field equations,  of course, depend on the choice made for the gravitational heat density $h_g$ of the null surface. Since we are varying $\ell^a$  in an extremum principle, it is natural to assume that $h_g$ is a quadratic in $\nabla_i\ell_j$ and hence will have the general form $\mathcal{P}^{abcd}\nabla_a\ell_c\nabla_b\ell_d$ where $\mathcal{P}^{abcd}$ is a tensor built from the background geometry. It can be shown that the variational principle will lead to a constraint on the background geometry only if $\mathcal{P}^{abcd}$ satisfies the conditions (b) and (c) mentioned earlier just after \eq{introP}. If we further assume  that either: (i) $\mathcal{P}^{abcd}$ is
built only from the metric or  (ii) we are in a $D=4$ spacetime, then the choice\cite{aseemtp1,aseemtp2,grtp} is \textit{unique} and  
$\mathcal{P}^{abcd}\propto P^{abcd}$ introduced in \eq{introP}. This will lead to the field equations in general relativity. (We will describe the more general case, when $\mathcal{P}^{abcd}$ can also depend on curvature, later on.) Thus, the variational principle is also entirely determined by the entropy tensor, and,
in general relativity, we will take $P^{ab}_{cd}=(1/2)(\delta^a_c\delta^b_d-\delta^b_c\delta^a_d)$ leading to:
\begin{equation}
h_g= -\frac{1}{4\pi} P^{ab}_{cd}\nabla_a\ell^c\nabla_b\ell^d
=\frac{1}{8\pi}[\nabla_a\ell^c\nabla_c\ell^a-(\nabla_a\ell^a)^2]
=-\frac{1}{8\pi} R_{ab}\ell^a\ell^b+{\rm (tot\ div)}
\label{introP1}
\end{equation}
This expression has another direct interpretation \cite{tpns1}. It can be shown that Einstein's equations, when projected on to \textit{any} null surface, take the form of a Navier-Stokes equation with the viscosity coefficients $\eta=1/16\pi,\zeta=-1/16\pi$ (This result was originally known for black hole horizons \cite{dns} and I generalized it to arbitrary null surfaces in Ref. \cite{tpns}.) Using this approach,  $h_g$ can be related\cite{tpns1} to the Navier-Stokes viscous tensor:
\begin{equation}
h_g= 2 \eta \sigma_{ab}\sigma^{ba} + \zeta \theta^2
\label{ns1}
\end{equation} 
where $\theta_{ab}\equiv q_a^iq_b^j\nabla_i\ell_j\equiv\sigma_{ab}+(1/2)q_{ab}\theta$ is the projection of $\nabla_i\ell_j$ on to the null surface and we have assumed that $\ell^a$ is affinely parametrized. This allows us interpret $h_g$ as the (fictitious\cite{tpns} but useful) viscous dissipation rate of the null surface which is being minimized in the extremum principle.
Another equivalent expression\cite{grtp} for the thermodynamic extremum principle, obtained by ignoring another  total divergence which does not contribute to the variation, 
 can be  based on the following integral over the null surface:
\begin{equation}
 Q\equiv \int_{\lambda_1}^{\lambda_2} d\lambda\ d^2x\, \sqrt{\sigma}\, \left[\frac{1}{16\pi}g^{ij} \ell_a\pounds_\ell N^a_{ij} +  T_{ab}\ell^a\ell^b \right]
\label{ns2}
\end{equation}
Extremising any of these expressions  
over all $\ell_a$ simultaneously will lead to the field equations. (The details of the derivation can be found in Refs. \cite{aseemtp1,aseemtp2,grtp}.) The result  will be in the form \eq{tf} if one uses the last expression ($R_{ab}\ell^a\ell^b$) in \eq{introP1} and will be in the form \cite{tpns1} of a Navier-Stokes equation if one uses the forms in \eq{ns1} or \eq{ns2}.

 Thus, a purely thermodynamic variational principle, invoking the extremization of the heat density of all the null surfaces simultaneously in the spacetime, leads to the gravitational field equation with an undetermined cosmological constant. We   see that the combination $g^{ij} \ell_a\pounds_\ell N^a_{ij}$ plays a vital role in the derivation of  the field equations from an extremum principle as well.

\subsection{Holographic interpretation of Einstein field equation}

The extremization of the heat density described above will lead to the field equations in the form $(2R_{ab} - T_{ab})\ell^a\ell_b =0$, which is equivalent to the standard Einstein's equation, with an undetermined cosmological constant appearing as an integration constant. But the field equation in this form has no simple physical meaning! \textit{If the gravitational field equations are the thermodynamic characterization of the spacetime, it should be possible to rewrite this equation in a  more transparent manner, in a thermodynamic language}. This is indeed possible\cite{grtp} using our  interpretation of Noether current; as a bonus we obtain an interesting concept involving the  bulk and surface degrees of freedom which I will call the holographic equipartition.\footnote{The term `holographic' is used here with its original meaning, indicating a surface-bulk correspondence. It has no (known) relation with the same adjective used in string theory in a different context.} 

To do this, 
we  will take the dot product of $u_a$ and the Noether current $J^a[\xi]$ in \eq{noe} (obtained with $v^a=\xi^a$), use \eq{fin1}, introduce the gravitational dynamics through $R_{ab} = (8\pi L_P^2)\bar T_{ab} $ (where $\bar T_{ab}\equiv T_{ab}-(1/2)g_{ab}T)$ and integrate the result over a 3-dimensional region $\mathcal{R}$
bounded by the equipotential surface.  Then we get \cite{grtp}:
\begin{equation}
 \frac{1}{8\pi L_P^2} \int_\mathcal{R} d^3x \sqrt{h}\, u_a g^{ij} \pounds_\xi N^a_{ij} =  \int_{\partial\mathcal{R}}\frac{d^2 x \, \sqrt{\sigma}}{L_P^2} 
\epsilon\left( \half k_B T_{\rm loc}\right) - \int_\mathcal{R}d^3x\, \sqrt{h}\, \rho_{\rm Komar} 
\label{holoevl}
\end{equation} 
where, in
the second term on the right hand side 
we have introduced the Komar energy density
  $2N\bar T_{ab} u^a u^b = (\rho + 3p)N$.

This result allows an interesting interpretation. If the spacetime is static and we choose the foliation such that $\xi^a$ is   the  Killing vector, then $\pounds_\xi N^a_{ij} =0$ and the left-hand-side vanishes. The equality of the two terms on the right-hand-side can be thought of as describing the \textit{holographic equipartition} \cite{grtp}, if we define the bulk and surface degrees of freedom along the following lines: The number of surface degrees of freedom is obtained by allotting one `bit' for each Planck area:
\begin{equation}
 N_{\rm sur}\equiv\frac{A}{L_P^2}=\int_{\partial \mathcal{R}} \frac{\sqrt{\sigma}\, d^2 x}{L_P^2}
\end{equation} 
 Further, we can define the \textit{average} temperature $T_{\rm avg}$ of the boundary surface $\partial\mathcal{R}$ to be:
 \begin{equation}
 T_{\rm avg}\equiv\frac{1}{A}\int_{\partial \mathcal{R}} \sqrt{\sigma}\, d^2 x\ T_{\rm loc}
\label{tav}
\end{equation} 
Finally, we will  define the number of  bulk degrees of freedom $N_{\rm bulk}$ by the following prescription: \textit{If} the matter in the region $\mathcal{R}$ is in equipartition at the average surface temperature $T_{\rm avg}$, \textit{then} we can identify $N_{\rm bulk}$ by $|E| = (1/2) N_{\rm bulk} k_B T_{\rm avg}$; that is, we define:
\begin{equation}
N_{\rm bulk}\equiv \frac{|E|}{(1/2)k_BT_{\rm avg}}= \frac{\epsilon}{(1/2)k_BT_{\rm avg}}\int_\mathcal{R} \sqrt{h}\ d^3x\; \rho_{\rm Komar}
\label{nbulkgen}
\end{equation} 
where $E$ is the total Komar energy in the bulk region $\mathcal{R}$ which is the source of gravity.
(The $\epsilon=\pm 1$  ensures that $N_{\rm bulk}$ remains positive even when the  Komar energy turns negative.)  
Our result in \eq{holoevl} implies that  \textit{comoving observers in any static spacetime} will indeed find:
\begin{equation}
 N_{\rm sur} = N_{\rm bulk} \qquad ({\rm Holographic \ equipartion})
 \label{key1}
\end{equation}
That is,  holographic equipartition holds  in all static spacetimes.

More importantly, \eq{holoevl} shows  that \textit{the departure from holographic equipartition --- leading to  a non-zero value for the right-hand-side ---  drives the dynamical evolution of the spacetime.} This is clear if we write \eq{holoevl} as:
\begin{equation}
\int \frac{d^3x}{8\pi L_P^2}\sqrt{h} u_a g^{ij} \pounds_\xi N^a_{ij} = \epsilon\frac{1}{2} k_B T_{\rm avg} ( N_{\rm sur} - N_{\rm bulk})
\label{evlnsnb}
\end{equation} 
One can also rewrite the left hand side of \eq{evlnsnb} by  relating $u_a g^{ij} \pounds_\xi N^a_{ij}$  to   more familiar constructs  using \eq{niskp}. This will give
\begin{equation}
 \int_\mathcal{V}\frac{d^3x}{8\pi L_P^2}h_{ab}\pounds_\xi p^{ab}  = \epsilon\frac{1}{2} k_B T_{\rm avg} ( N_{\rm bulk} - N_{\rm sur})
\label{start}
\end{equation} 
which allows us to connect the thermodynamic interpretation with the standard  Hamiltonian formulation of relativity.
Demanding the validity of \eq{start} or \eq{evlnsnb} for all foliations
 is equivalent to demanding the validity of 
Einstein's equations.

Even in a static spacetime, non-static observers  will perceive a departure from holographic equipartition because \eq{evlnsnb}  --- while being generally covariant --- is dependent on the foliation  through the normal $u_i$.   A natural foliation which we can use in \textit{any} (local region of) spacetime is the synchronous frame in which $u_a$ will be the velocity of geodesic observers. In this case, the acceleration (and thus the temperature and the Noether potential) vanishes and \eq{evlnsnb} can be replaced by the local vector equation:
\begin{equation}\label{Paper1:Eq21}
g^{ij}\pounds _{\xi}N^{a}_{ij}+ \bar T^a_{b}u^b\equiv \bar P^a+ \bar T^a_{b}u^b=0;
\quad \bar P^a\equiv g^{ij}\pounds _{\xi}N^{a}_{ij}
\end{equation}
Since $\bar T^a_{b}u^b$  is the flux of the Komar energy (defined using $\bar T^a_b\equiv T^a_b-(1/2)\delta^a_bT$), this result can be interpreted as a momentum balance equation between the gravitational momentum and the matter momentum fluxes. This provides a simple interpretation of 
the, by now familiar, combination $\bar P^a=g^{ij}\pounds _{\xi}N^{a}_{ij}$ as measured by \textit{geodesic observers}.

Taking a cue from this, one can obtain a more general result \textit{valid for any spacetime (and foliation)}, which could be time-dependent and dynamically evolving. The total energy contained in a region $\mathcal{R}$ bounded by an equipotential surface $\partial\mathcal{R}$
is  exactly equal to the surface heat content, when the equations of motion hold:
  \begin{equation}
 \int_\mathcal{R} d^3x\sqrt{h}u_a[\bar P^a(\xi)+N\bar T^a_bu^b]=\epsilon\int_{\partial\mathcal{R}} d^2x \ Ts
\label{energytotal}
\end{equation}
  In the synchronous frame, the acceleration temperature on the right hand side of \eq{energytotal} vanishes and so does the left hand side (see \eq{Paper1:Eq21}); in any other, arbitrary foliation, the effect of acceleration is captured by the surface term on the right hand side.
If one prefers to use $T_{ab}$ rather than $\bar T_{ab}$, that can be easily done by using a closely related gravitational momentum vector, defined by:
\begin{equation}
P^a\equiv J^a[\xi]-2G^a_b \xi^b=g^{ik}\pounds_\xi N^{a}_{ik}+R\xi^a\equiv \bar P^a+R\xi^a
\label{defpa1}
\end{equation}
so that $\bar P^a(\xi)+N\bar T^a_bu^b=P^a(\xi)+NT^a_bu^b$ on-shell 
and either of them can be used in the left hand side of \eq{energytotal}.
 (For a more detailed discussion of these results, and properties of $P^a$, see Ref. \cite{grtp}).

\section{The value of the Cosmological Constant}

Once we accept that gravitational field equations are invariant under $T^a_b \to T^a_b + ({\rm constant})\ \delta^a_b$, the solution will have an undetermined cosmological constant arising as integration constant. We then need a \textit{new} physical principle to determine its value which I will now describe \cite{hptp1,hptp2}.

Observations indicate that our universe is
  characterized by (i) an early inflationary phase with approximately constant density $\rho_{inf}$; (ii) a phase dominated by radiation and matter, with $\rho=\rho_{eq}[x^{-4}+x^{-3}]$ where $x(t)\equiv a(t)/a_{\rm eq}$ and $\rho_{eq}$ is another constant and $a_{eq}$ is the epoch at which matter and radiation densities were equal; and (iii) an accelerated phase of expansion at late time dominated by the energy of the cosmological constant $\rho_\Lambda$. Thus, there are three undetermined densities $[\rho_{inf},\rho_{eq},\rho_\Lambda]$ which will completely describe the dynamics of our universe. It is generally believed that high energy physics will eventually determine $\rho_{inf}$ and $\rho_{eq}$ but we need a new principle to fix the value of $\rho_\Lambda$.

It turns out that, a universe with these three phases has a \textit{conserved} quantity, viz. the number $N$ of length scales which cross the Hubble radius during each of these phases. It can be shown that  $N(a_2,a_1)= (2/3\pi)\ln(H_2a_2/H_1a_1)$ during any interval $a_1<a<a_2$. Any physical principle which fixes the value of $N$ during the radiation-matter dominated phase, say, will relate $\rho_\Lambda$ to $[\rho_{inf},\rho_{eq}]$. We have given arguments elsewhere \cite{hptp1,hptp2} as to why we expect $N=4\pi$ which leads to the remarkable relation connecting the three densities:
\begin{equation}
 \rho_\Lambda  \approx \frac{4}{27} \frac{\rho_{\rm inf}^{3/2}}{\rho_{\rm eq}^{1/2}} \exp (- 36\pi^2)
\label{ll6}
 \end{equation} 
For the observed range of $\rho_{eq}$, and the range of inflationary energy scale  $\rho_{\rm inf}^{1/4} = (1.084-1.241) \times 10^{15}$ GeV, we get $\rho_{\Lambda} L_P^4  = (1.204 - 1.500) \times 10^{-123}$, which is consistent with observational results! I will confine myself to just two brief comments about this result  here; more details can be found in Ref. \cite{hptp1,hptp2}.

First, this is a very novel approach for solving the cosmological constant problem based on a unified view of cosmic evolution, connecting all the three phases through \eq{ll6}. This is in contrast to standard cosmology where the three phases are put together in an unrelated, ad hoc, manner.

Second, it is difficult to incorporate $N=4\pi$ into the standard cosmological paradigm. But it fits naturally into the concept of holographic equipartition. It turns out that the Friedmann equation itself can be written in a very suggestive  form as:
\begin{equation}
\frac{dV}{dt} = L_P^2 (N_{\rm sur} - \epsilon N_{\rm bulk}); 
\end{equation} 
with $V= (4\pi/3H^3)$ and  $N_{\rm sur}, N_{\rm bulk}$ defined as before with the Unruh-Davies temperature for the horizon being taken as $T=H/2\pi$. 
In this approach, the value of the conserved quantity $N$ gets fixed at the Planck scale as $N=4\pi L_P^2/L_P^2 = 4\pi$. In such a model, the assumption $N=4\pi$ can arise very naturally.

\section{Thermodynamic description of \LL\ models}

In standard thermodynamics, one can use any suitable thermodynamic functional like entropy ($s$), heat density ($Ts$), free energy ($\rho-Ts$) etc., to describe the state of matter. Similarly, the gravitational field equations, which describe the state of the spacetime, are encoded in the thermodynamic potential for the spacetime.

In the case of general relativity, the Noether potential associated with the time evolution vector $\xi^j$ and
the entropy density of a patch of the local Rindler horizon are governed by the tensor $P^{abcd}$ used in \eq{introP} which satisfies the three conditions:  (a) it is made from the metric tensor; (b) it has all the algebraic symmetries of the curvature tensor; (c) it is divergence free in all indices. 
Further,  the thermodynamic variational principle used in general relativity, to obtain the field equations, has the gravitational part given by $P^{ab}_{cd}\nabla_a\ell^c\nabla_b\ell^d$ (see \eq{introP1} and thus the field equations can be obtained 
once we know the entropy density tensor $P^{ab}_{cd}$.

Since the dynamics of spacetime can be related to the form of $P^{abcd}$, more general theories of gravity can be obtained by relaxing the assumptions (a), (b) or (c) above. It can be shown that if the field equations have to be of second order in the metric, then the \textit{only} assumption we can drop is (a) in the list. If we 
allow  the entropy density tensor $P^{abcd}$ to depend on both metric and curvature (while obeying the conditions (b) and (c)) we obtain a more general class of theories of gravity called the \LL\ models. (Incidentally, in $D=4$, the \LL\ model reduces back to general relativity but for $D>4$  it is different). The form of $P^{abcd}$ in these theories are uniquely   determined\cite{llreview,ll} by conditions (b) and (c) and is given by: 
\begin{equation}\label{Paper2:Eq11}
P^{ab}_{cd}
=m\delta ^{aba_{2}b_{2}\ldots a_{m}b_{m}}_{cdc_{2}d_{2}\ldots c_{m}d_{m}}
R^{c_{2}d_{2}}_{a_{2}b_{2}}\ldots R^{c_{m}d_{m}}_{a_{m}b_{m}}
\end{equation}
The heat density  of the null horizons is now given by the integral 
\begin{equation}
\mathcal{S}=-\frac{1}{8}\int d^{D-2}x \sqrt{\sigma}P^{abcd}\epsilon_{ab}\epsilon_{cd}\equiv\int d^{D-2}x\ s
\label{sctp56}
\end{equation} 
The Noether potential associated with $\xi^a$ has the same form as before (see \eq{introP}) with the corresponding $P^{abcd}$ and is given by:
$16 \pi J^{ab}\left(\xi \right) = 2P^{abcd}\nabla _{c}\xi _{d}$,
so that the  Noether current is:
\begin{equation}
16 \pi J^{a} = 2\mathcal{R}^{a}_{b}\xi ^{b}+P_{i}^{~jka}\pounds _{\xi}\Gamma ^{i}_{jk}; \qquad
\mathcal{R}^{ab}\equiv P^{aijk}R^{b}_{~ijk}.
\end{equation} 
Given these constructs, all the previous thermodynamic interpretation go through in a natural fashion and one can generalize\cite{sumantatp} all the previous results to \LL\ models along the following lines. 

\begin{itemize}

 \item 
The Noether charge density in this case is given by
\begin{equation}
16 \pi u_{a}J^{a}(\xi)=2D_{\alpha}\left(N\chi ^{\alpha}\right); \qquad
\chi ^{a} \equiv 2P^{abcd}u_{b}u_{c}a_{d}
\end{equation}
This vector $\chi^a$ generalizes the notion of acceleration in this case. Integrating this result, it is straightforward to show that \cite{sumantatp} 
\begin{equation}
\int _{\mathcal{V}}d^{D-1}x \sqrt{h} u^{a}J_{a}\left(\xi \right)
=\epsilon \int _{\partial \mathcal{V}}d^{D-2}x~T_{loc}s.
\end{equation}
where $s$ is defined by \eq{sctp56}.
Thus  \emph{the 
Noether charge in a bulk region is equal to the  heat content of the boundary in all \LL models as well }.

\item 
The field equations of \LL\ models can be obtained \cite{aseemtp1,aseemtp2} by extremizing a heat density just as before (see \eq{introP1}) with a gravitational contribution proportional to $P^{abcd}\nabla_a \ell_c \nabla_b\ell_d$. This leads to the \LL\ field equations with the cosmological constant arising as an integration constant.
\item
More importantly, the ideas of holographic equipartition work out seamlessly in all \LL\ models.  One can show \cite{sumantatp} that the field equations can be written in the form 
\begin{equation}
\frac{1}{8\pi}\int _{\mathcal{R}}d^{d-1}x\sqrt{h}(2u_{a}P_{i}^{~jka}\pounds _{\xi}\Gamma ^{i}_{jk})
=\epsilon \left(\frac{1}{2}T_{avg}\right)\left(N_{sur}-N_{bulk}\right)
\end{equation}
where $N_{\rm sur}, T_{\rm avg}$ and $N_{\rm bulk}$ are defined by direct generalization \cite{sumantatp} of the corresponding expressions in the case of general relativity:
\end{itemize}

The fact that one could generalize the thermodynamic interpretation to all \LL\ models is quite nontrivial because, in a general \LL\ model, the entropy is \textit{not} proportional to the area. 
This suggests the thermodynamic perspective encodes some deeper feature about the spacetime not captured within general relativity and will possibly be revealed only when we have a complete description of quantum microstructure.

\section{Heat density of spacetime from the zero-point length}

We have seen above that the gravitational field equations in general relativity can be obtained \cite{llreview} by extremising the total heat density  $\mathcal{S}=\mathcal{S}_g+\mathcal{S}_m$ where the  gravitational heat density $\mathcal{S}_g[\ell]$  is given by \cite{grtp} 
\begin{equation}
 \mathcal{S}_g\propto [(\nabla_i\ell^i)^2-\nabla_i \ell^j \nabla_j\ell^i]=R_{ab} \ell^a\ell^b +(\text{tot.  div.})
\label{ts}
\end{equation}
 If the ideas of the emergent gravity paradigm are correct,
it must be possible
  to obtain this expression from a more microscopic approach. I will now show how this can be done\cite{cheshire1,cheshire2}.
 
To begin with, the existence of some `atoms of spacetime' is related to an effective \textit{discreteness} at Planck scale ($L_P^2 = (G\hbar / c^3)$), which allows us to assign  $N_{\rm sur}=A_{\perp}/L_P^2$ degrees of freedom with any area $A_{\perp}$.  So we need to incorporate the notion of `zero-point-area' $L_P^2$ in a suitable manner if we hope to obtain $\mathcal{S}_g $ from a more fundamental description. 
Further, we saw earlier that  the  entropy \textit{density} of spacetime, when evaluated around any event after Euclidean continuation, is  given by $(K\sqrt{h})$.
These two facts suggest that one should be able to obtain the entropy density in \eq{ts} from $K\sqrt{h}$ in a suitable limit, \textit{if we can introduce the zero-point-length in to the computation} of $K\sqrt{h}$ consistently.

The operational difficulty in implementing this idea, of course, is the following: Expressions  like $K\sqrt{h}$ are well defined on a differentiable manifold with a metric, normal vectors to foliations  etc. But the entropy arising from  ($A_\perp/L_P^2$)  degrees of freedom associated with an area $A_\perp$, requires introducing  the  zero point area into the spacetime. This cannot be done without modifying the usual, local, description of spacetime. 

We need a  prescription which incorporates the quantum gravitational effects (in particular the existence of zero point area $L_P^2$), at scales reasonably bigger than $L_P^2$ but not completely classical.
 This will require the concept of an ``effective'' metric $q_{ab}$ in a spacetime (which has  classical metric $g_{ab}$) such that it can incorporate the  effects of the zero point area $L_P^2$. If we compute $K\sqrt{h}$ for this effective metric, then in the  appropriate limit, this should give us the entropy density of the spacetime. Further, if our ideas are correct, the resulting entropy density should match with the one in \eq{ts}.
 Fortunately,  this key  step of introducing an effective metric $q_{ab}$ with the necessary properties is already done in Ref. \cite{cheshire1,cheshire2}.
I will briefly describe this procedure.

In any classical spacetime, one can introduce a geodesic interval $\sigma^2(P,p)$ between any two events $P$ and $p$ which carries the same amount of information as the metric  $g_{ab}$.
The key difference is that $\sigma^2(P,p)$ is a biscalar (and hence nonlocal) while the metric tensor is local. Geometric quantities at $P$ can be obtained by taking suitable derivatives of $\sigma^2(P,p)$ with respect to the coordinate $p$ and then taking the limit $p\to P$. (See Ref.~\cite{cheshire1,cheshire2,poisson-lrr}.) Classical geometry can be characterized either by $g_{ab}$ or by $\sigma^2(P,p)$.

When we attempt to incorporate the effects of quantum gravity, there is a distinct advantage in starting from a description in terms of $\sigma^2(P,p)$ rather than from the metric. This is because we have no universal rule to describe
how quantum gravity modifies the metric; but  there is a significant amount of evidence (see e.g., Ref.~\cite{zpl}) which suggests that $\sigma^2(P,p)$ is modified by
\begin{equation}
 \sigma^2 \to \sigma^2 + \lp^2; \qquad \lp^2=\mu^2 L_P^2
\label{funda}
\end{equation}
where $\mu$ is a factor of order unity
\footnote{As an aside, we mention that, it is possible to consider the general case in which $\sigma^2 \to S(\sigma^2)$ where the function $S$ will  arise from a more fundamental framework 
for quantum gravity. Surprisingly, our  result turns out to be independent of the  form for $S(\sigma^2)$ as long as $S(0)=L_0^2$!}. That is, one can capture the lowest order quantum gravitational effects by introducing a zero point length in spacetime through the prescription in \eq{funda}. 

We can now define a second rank symmetric \textit{bitensor} $q_{ab}(p,P)$, (called `qmetric') constructed such that it will have the geodesic interval $\sigma^2 + L_0^2$ if the original metric had the geodesic interval $\sigma^2$. This requires associating with  a metric $g_{ab}$ (which has the  geodesic interval $\sigma^2$) a \textit{nonlocal} symmetric bitensor $q_{ab}(p,P)$ by the relation:  
 \begin{equation}
q_{ab}(p,P; \lp^2) \equiv Ag_{ab} -    \l( A  - \frac{1}{A} \r) \; \ell_a \ell_b
\label{eq:key1}
\end{equation}
where $g_{ab} = g_{ab}(p)$  is the classical metric tensor, $\sigma^2 = \sigma^2 (p,P)$ is the corresponding classical geodesic interval  and 
\begin{equation}
\mA \equiv 1 + \frac{\lp^2}{\sigma^2}; \qquad \ell_a = \frac{\nabla_a \sigma^2}{2 \sqrt{  \sigma^2}}
\label{n1}
\end{equation}
The qmetric  can capture some of the effects of quantum gravity --- especially those arising from the existence of the zero point length --- \textit{without us leaving the comforts of the standard differential geometry.} 

The key reason why several non-trivial effects arise from such a nonlocal description of geometry in terms of the qmetric is  
 the following: Suppose $\phi(P|g)$ is a scalar computed from the metric $g_{ab}$ and its derivatives (e.g., $\phi$ could be $R$, or $R_{ab}R^{ab}$ etc.). When we carry out the corresponding algebra using $q_{ab}(p,P)$ instead of $g_{ab}(p)$ (with  differentiations carried out at the event $p$) we will  get a nonlocal (biscalar) $\phi(p,P; \lp^2|q)$ which will depend on two events $(p,P)$ and on $\lp^2$. To get a local result, we  take the limit of $\sigma\to0$ (that is, $p\to P$) keeping $L_0^2$ finite. \textit{The resulting form of  $\phi(P,P; \lp^2|q)$ will exhibit quantum gravitational residual effects due to nonzero $L_0^2$. }
This  arises from the non-commutativity of the limits:
\begin{equation}
\lim_{L_0^2\to0}\lim_{\sigma^2\to0} \phi(p,P; \lp^2|q)\neq
\lim_{\sigma^2\to0}\lim_{L_0^2\to0}\phi(p,P; \lp^2|q)
\label{ncl}
\end{equation} 
The limit on the right hand side of this equation is trivial.
When we  take the limit of $L_0^2\to0$, keeping $\sigma^2$ finite, we find that $q_{ab}\to g_{ab}$ and  $\phi(p,P; \lp^2|q)\to\phi(P|g)$. This is because, in \eq{eq:key1}, only the combination $L_0^2/\sigma^2$ introduces any non-trivial effects and this term (and all related derivatives) will vanish when $L_0^2\to0$. But if we take the limit of $\sigma\to0$ (with finite $L_0^2$), the qmetric actually diverges. Therefore, we have no assurance that we will even get anything sensible when we take the limit in the left hand side of \eq{ncl}; surprisingly, we do, leading to   non-trivial effects. 

I will  now briefly describe the results of this analysis working in a $D=4$ Euclidean space  and using units with $L_P=1$ so that $L_0=\mu$. Given a  spacetime event $P$, the most natural {\it surface} $\Sigma$ on which to evaluate $K\sqrt{h}$ is the one formed by events $p$ at a constant  geodesic interval $\sigma(p,P)=\lambda$ from $P$. In the local Rindler frame around $P$, the origin of the $t_E-x$ plane will be the local Rindler horizon;   the limit of $p\to P$ corresponds to computing a quantity on the horizon. Computing $K\sqrt{h}(p,P,\mu^2)$ using the qmetric and taking the limit $p\to P$  to get the quantum corrected entropy density is straightforward, and we get:
\begin{eqnarray}
K\sqrt{h} = 3 \mu^2 - \frac{\mu^4}{3} \ R_{ab} \ell^a\ell^b 
= \mathcal{S}_0 - \frac{\mu^4}{3} \mathcal{S}_g
\label{final1}
 \end{eqnarray}
with all quantities  now evaluated at $P$. 
The term $\mathcal{S}_0 = 3\mu^2$ can be thought of as the zero point entropy density of the spacetime which is a new feature. Its numerical value depends on the ratio $\mu=L_0/L_P$ which we expect to be of order unity (We will comment more about this later on). \textit{The second term in \eq{final1} is exactly the heat density used in the emergent gravity paradigm.}
 This result is significant in several ways which I will now describe.
 
 The most important feature, of course,  is that it reproduces correctly (except for an unimportant multiplicative constant) the entropy density $\mathcal{S}_g \propto R_{ab} \ell^a \ell^b$ we used in our approach. Further, the negative sign of the second term in \eq{final1} is important for the consistency of this result. \textit{These facts show that the entire program has a remarkable level of internal consistency.} One could not have guessed the limit a priori and, in fact, there is no assurance that the limit should even be finite in the coincidence limit of $\sigma^2 \to 0$.   It is  a non-trivial feature that the final result is free of any divergence.\footnote{Incidentally, if we attempted the same computation with the bulk cosmological constant term in the action ($\Lambda\sqrt{-g}$) by replacing it by ($\Lambda\sqrt{-q}$)
and taking the $\sigma^2\to0$ limit, it will diverge unless $\Lambda=0$. Probably this result indicates that a microscopic approach cannot accommodate a nonzero cosmological constant.} 

 As we mentioned earlier, $K\sqrt{h}$ has the  interpretation of (being proportional to) the heat density on the horizon. After analytic continuation, in the Euclideanized local Rindler frame around an event $P$, the Rindler horizon gets mapped to the origin of the $(x, t_E)$ plane. The coincidence limit of $p\to P$ is then the same as taking the horizon limit. In this limit,  $K\sqrt{h}/8\pi$ gives the entropy density (see \eq{entro}). So if we had taken the limit $L_0\to0$  first (so that $q_{ab}\to g_{ab}$ etc.) we would have been led to this standard result. We get the nontrivial result in \eq{final1} only because of taking limits in the appropriate manner.

Finally, an intriguing aspect of our result is the discovery of the ``zero point entropy density'' represented by the first term
 $\mathcal{S}_0 = 3\mu^2$ in \eq{final1}. The total zero point entropy in a sphere of Planck radius is given by
 \begin{equation}
  S_0 = \frac{4\pi}{3} \times 3 \mu^2 = 4\pi\mu^2
  \label{fourpi}
 \end{equation} 
We saw earlier that the cosmological constant problem can be solved within the emergent gravity paradigm if one could attribute a value $4\pi$ to the measure of degrees of freedom in the universe for a GUT scale inflation.  On the other hand, if inflation took place at Planck scales, we need $\mu^2\approx 1.2$ (see Ref.~\cite{tpqg})
all of which is quite consistent with \eq{fourpi}.
Unfortunately, the value of $\mu$ cannot be determined from the analysis of the pure gravity sector, but the fact that we get a non-zero residual entropy density is encouraging.   

\section*{Acknowledgments}
I thank Sumanta Chakraborty, Dawood Kothawala, Bibhas Majhi and Krishnamohan Parattu for discussions. My work is partially supported by the J.C.Bose research grant of the Department of Science and Technology, Government of India.


\vskip 0.4in

\begin{thebibliography}{99}

\bibitem{du} 
P. C. W. Davies,  \textit{J. Phys.}, \textbf{A 8},  609 (1975). 

W. G. Unruh,   \textit{Phys. Rev.} \textbf{D 14}, 870 (1976).

\bibitem{KBP} Krishnamohan Parattu, Bibhas Ranjan Majhi, T. Padmanabhan,    \textit{Phys.Rev.,} \textbf{D 87},  124011 (2013) [arXiv:1303.1535].

\bibitem{grtp}  
T. Padmanabhan,  Gen.Rel.Grav, \textbf{46}, 1673 (2014) [arXiv:1312.3253].


\bibitem{grtp28} 
 J. Katz,     \textit{Class. Quantum Grav.}, \textbf{2}, 423 (1985).


\bibitem{aseemtp1} 
T. Padmanabhan and Aseem Paranjape,  \textit{Phys.Rev.} \textbf{D75}, 064004, (2007) [gr-qc/0701003]

\bibitem{aseemtp2} 
 T. Padmanabhan,  \textit{Gen.Rel.Grav.}, \textbf{40}, 529 (2008) [arXiv:0705.2533] 


\bibitem{tpap}
T. Padmanabhan,   \textit{Adv. Sci. Lett.}, \textbf{2}, 174 (2009) [arXiv:0807.2356] 

\bibitem{hptp1}
Hamsa Padmanabhan and  T. Padmanabhan,  \textit{Int.Jour.Mod.Phys,} \textbf{D22}, 1342001 (2013) [arXiv:1302.3226].

\bibitem{hptp2} 
T. Padmanabhan and Hamsa Padmanabhan, \textit{Int.Jour.Mod. Phys.}, \textbf{D 23}, (2014)  1430011 [arXiv:1404.2284]. 

\bibitem{llreview} 
T. Padmanabhan and Dawood Kothawala,  {Phys.Repts.} \textbf{531},115(2013) [arXiv:1302.2151].

\bibitem{cheshire1} 
Dawood Kothawala and  T. Padmanabhan (2014),  \textit{Grin of the Cheshire cat: Entropy density of spacetime as a relic from quantum gravity} [arXiv:1405.4967].

\bibitem{cheshire2} 
Dawood Kothawala and T. Padmanabhan (2014), \textit{Entropy density of spacetime from the zero point length,} [arXiv:1408.3963]

\bibitem{zpl} 
B. S. DeWitt,  (1964) Phys.\ Rev.\ Lett.\ {\bf 13}, 114;
T. Padmanabhan, (1985) Gen. Rel. Grav.\ {\bf 17} , 215 ;  (1985) Ann. Phys.\ {\bf 165}, 38.
For a review, see L. Garay, (1995) Int.\ J.\ Mod.\ Phys.\ A {\bf 10}, 145 .

\bibitem{carlip} S. Carlip, \textit{Challenges for Emergent Gravity}, [arXiv:1207.2504].

\bibitem{insights}  T. Padmanabhan,  \textit{Rept. Prog. Phys.}, \textbf{ 73} (2010) 046901, [arXiv:0911.5004]

\bibitem{a12} T. Jacobson, (1995) \textit{Phys.Rev.Lett.} \textbf{75} 1260, [gr-qc/9504004]. 

\bibitem{a41} B. R. Majhi and T. Padmanabhan, (2013) \textit{Eur. Phys. J.} C , \textbf{73}:2651, [arXiv:1302.1206].


\bibitem{unimod1}
Ellis, G. F. R. (2013), \textit{The Trace-Free Einstein Equations and inflation}, [arXiv:1306.3021];

\bibitem{unimod3}
David R. Finkelstein, Andrei A. Galiautdinov and James E. Baugh (2001),  \textit{J. Math. Phys.} \textbf{42}, 340 [gr-qc/0009099];

\bibitem{unimod4}
William G. Unruh, (1989)  \textit{Phys. Rev.} \textbf{D 40} 1048.

\bibitem{tpns1} Sanved Kolekar, T. Padmanabhan,  \textit{Phys.Rev.,} \textbf{D 85}, 024004 (2012) [arXiv:1109.5353].

\bibitem{dns}
T. Damour, (1982), \textit{Surface effects in black hole physics},
in \textit{Proceedings of the Second Marcel Grossmann Meeting on General
Relativity}, Ed. R. Ruffini, North Holland , p. 587;
Thorne K S et al. 1986 \textit{Black Holes: The Membrane Paradigm} (Yale University Press, 1986)
  
\bibitem{tpns}
T. Padmanabhan,  \textit{Phys.Rev.,} \textbf{D 83}, 044048 (2011) [arXiv:1012.0119] 



\bibitem{sumantatp} 
Sumanta Chakraborty, T. Padmanabhan, \textit{Evolution of Spacetime arises due to the departure from Holographic Equipartition in all Lanczos-Lovelock Theories of Gravity}, [arXiv:1408.4679];
 
Sumanta Chakraborty, T. Padmanabhan,  \textit{Phys.Rev.},  \textbf{D 90}, 084021 (2014)
 [arXiv:1408.4791].

\bibitem{ll}C. Lanczos, \textit{Z.Phys.} \textbf{73}, 147 (1932);
D. Lovelock, \textit{J. Math. Phys.} \textbf{12}, 498 (1971).

\bibitem{poisson-lrr}
E.~Poisson, A.~Pound, I.~Vega,  (2011) Liv.\ Rev.\ Rel.\ {\bf 14}, 7.

\bibitem{tpqg}
T. Padmanabhan, \textit{The Physical Principle that determines the Value of the Cosmological Constant}, [arXiv:1210.4174].

\end{thebibliography}
\end{document}